\documentclass[letter,dvipdfmx]{jpsj3}
\usepackage[dvipdfmx]{graphicx}
\usepackage[dvipdfmx]{color}

\renewcommand{\a}{\alpha}
\renewcommand{\b}{\beta}
\newcommand{\bea}{\begin{eqnarray}}
\newcommand{\eea}{\end{eqnarray}}
\newcommand{\f}[2]{\frac{#1}{#2}}
\newcommand{\eq}{&=&}
\newcommand{\nn}{\nonumber \\ }

\newcommand{\area}{\int_{-\infty}^\infty }

\renewcommand{\l}{\lambda}
\newcommand{\p}{\partial}
\newcommand{\pp}[2]{\f{\p #1}{\p #2}}

\newcommand{\siki}[1]{Eq. (\ref{#1})}
\newcommand{\sikis}[2]{Eqs. (\ref{#1}) and (\ref{#2})}

\allowdisplaybreaks[1]

\newcommand{\g}{\gamma}
\renewcommand{\l}{\lambda}
\allowdisplaybreaks[1]

\title{
Replica Analysis for Maximization of Net Present Value
}

\author{Takashi Shinzato\thanks{shinzato@eng.tamagawa.ac.jp}
}
\inst{
Department of Management Science, College of Engineering, 
Tamagawa University, \\ Machida, Tokyo 1948610, Japan\\
\smallskip
{\rm (Received October 14, 2018; accepted *** 1, 2018)}
} %\\

\abst{
In this paper, we use replica analysis to determine the investment strategy that {can maximize} the net present value for portfolios containing multiple development projects. 
Replica analysis was developed in statistical mechanical informatics and econophysics to evaluate {disordered} systems, and here we use it to formulate the maximization of the net present value as an optimization problem under budget and investment concentration constraints.
 Furthermore, 
{we confirm that a common approach from operations research underestimates the true {maximal} net present value as the {maximal} expected net present value by comparing our results with the maximal expected net present value as derived in operations research.} Moreover, it is shown that the conventional method for estimating the net present value does not consider variance in the cash flow.
}
\begin{document}
\maketitle

%\section{Introduction}
In recent decades, 
{in order to grow their businesses,
companies have used indicators based on internal interest rates, net present value, yield to redemption, etc., and these indicators have aided} decision making, particularly with respect to development projects, for example, real estate development by real estate companies and the development of new drugs by pharmaceutical companies
\cite{Luenberger1998InvestmentScience,marcus2014investments}.
Meanwhile, in mathematical finance research, there have been several attempts to evaluate the investment value of a project while taking into consideration uncertainty regarding the cash flow generated by the project during the investment period. Estimating the expected investment value from multiple projects by the maximization of the expected utility and determining how to diversify portfolios constitute an active area of research\cite{WIESEMANN2010356,
doi:10.1287/mnsc.16.5.357,
YANG1995327,
ELMAGHRABY199035,
doi:10.1287/mnsc.23.8.882,
SMITHDANIELS198733}.
It is difficult to determine the minimal investment risk and maximal expected return with the conventional operations research approach, {which is related to analyzing annealed disordered systems in the context of spin glass theory.} Thus, the portfolio optimization problem has been analyzed using analytical methods from statistical mechanical informatics and econophysics, and the findings compared with those of conventional methods from operations research\cite{
1742-5468-2017-12-123402,
1742-5468-2016-12-123404,
doi:10.1080/1351847X.2011.601661,
KONDOR20071545,
PAFKA2003487,
Pafka2002,
Ciliberti2007,
doi:10.1080/14697680701422089,
110008689817,
doi:10.7566/JPSJ.86.063802,
doi:10.7566/JPSJ.86.124804,
SHINZATO2018986,
PhysRevE.94.052307,
PhysRevE.94.062102b,
10.1371/journal.pone.0134968,
10.1371/journal.pone.0133846,
1742-5468-2017-2-023301,
Ryosuke-Wakai2014}.
Indicators for minimizing the investment risk, maximizing the expected return, and reducing the investment concentration in portfolio optimization have been analyzed, but there has been little research conducted on evaluating the optimal solution for 
{decision making related to multiple investment projects.}
The maximization of the net present value in investment projects is a stochastic optimization problem, and methods for quenched disordered systems from spin glass theory literature have been used to analyze the optimal investment strategy{\cite{
110008689817,
doi:10.7566/JPSJ.86.063802,
doi:10.7566/JPSJ.86.124804,
SHINZATO2018986,
PhysRevE.94.052307,
PhysRevE.94.062102b,
10.1371/journal.pone.0134968,
10.1371/journal.pone.0133846,
1742-5468-2017-2-023301,
Ryosuke-Wakai2014}.}
In this study, methods for quenched disordered systems are used to obtain the {maximal} net present value, under budget and investment concentration constraints, and the investment allocation for maximizing the sum of the net present value over multiple investment projects. The maximal net present value will also be considered using replica analysis, which is a powerful analysis method from statistical mechanical informatics.

%\section{Model Setting}

Here, we consider a portfolio of investments in development projects, and seek to maximize the total net present value for $N$ investment projects (hereinafter referred to as projects), such as real estate development and drug development, under constant conditions.
The present period ($t=0$) is taken as the 
beginning of the investment period,
the amount invested into project $i(=1,2,\cdots,N)$
is $w_i$, the amounts invested across all projects are represented as $\vec{w}=(w_1,w_2,\cdots,w_N)^{\rm T}\in{\bf R}^N$ 
(the total amount invested is the sum $\sum_{i=1}^Nw_i=Nm$, where $m$ is the initial budget of each project),and $r(>0)$ is the interest rate. 
Each project $i$ generates a cash flow until it is finished at maturity period $t=T$, where the notation ${\rm T}$ indicates 
the transpose of a vector or matrix.
Moreover, 
the divestment amount of project $i$ at maturity period $T$ is 
$\l_iw_i$ 
with attenuation rate $\l_i\ge0$.
Then, the net present value of project $i$
${\rm NPV}_i$ is described as 
\bea
\label{eq1}
{\rm NPV}_i=
-w_i+
\sum_{t=1}^T\f{c_iw_i+c_iw_ix_{it}}{(1+r)^t}
+\f{\l_iw_i}{(1+r)^T},
\eea
where $c_i\ge0$ is the coupon rate of project $i$, 
and the cash flow in each investment period is 
$c_iw_i+c_iw_ix_{it}$, in which $c_iw_i$ is the mean cash flow and $c_iw_ix_{it}$ represents random fluctuations in the cash flow. Moreover, we assume that the random element $x_{it}$
is independent and identically distributed (i.i.d.) with mean 
and variance $E[x_{it}]=0$ and $V[x_{it}]=v_i$, respectively. From this, 
the sum of the net present value of $N$ projects 
for investment amount $\vec{w}$
is defined by
\bea
\label{eq2}
{\cal H}(\vec{w}|X)
\eq
\sum_{i=1}^N{\rm NPV}_i,
\eea
where $X=\left\{x_{it}\right\}\in{\bf R}^{N\times T}$ is a matrix representing random fluctuations in the cash flow $x_{it}$. Furthermore, 
the budget constraint (\siki{eq3}) and the investment concentration constraint (\siki{eq4}) with respect to the investment amount $\vec{w}$ are defined by 
\bea
\label{eq3}\sum_{i=1}^Nw_i\eq Nm,\\
\label{eq4}\sum_{i=1}^Nw_i^2&\le& N\tau,
\eea
where \siki{eq4} represents the expansion of the 
Herfindahl-Hirschman index, that is, the investment concentration $q_w=\f{1}{N}\sum_{i=1}^Nw_i^2$.
Moreover, regarding the budget and investment concentration constraints, {since $\f{1}{N}\sum_{i=1}^Nw_i^2-\left(\f{1}{N}\sum_{i=1}^Nw_i\right)^2=
\f{1}{N}\sum_{i=1}^N
\left(w_i-\f{1}{N}\sum_{i=1}^Nw_i\right)^2
\ge0$ holds, $\tau\ge m^2$ can be easily obtained.}

For investment portfolios,
{we wish to determine the portfolio which can maximize} the net present value ${\cal H}(\vec{w}|X)$ in \siki{eq2} 
with respect to the investment amount $\vec{w}$ under the constraints of \sikis{eq3}{eq4}.
From the above definitions, 
the maximal net present value for a project 
is defined as
\bea
{\kappa}\eq\f{1}{N}
\mathop{\max}_{\vec{w}\in{\cal D}}{\cal H}(\vec{w}|X)
,
\eea
where 
the constraints in Eqs. (\ref{eq3}) and 
(\ref{eq4}) with respect to the investment amount 
$\vec{w}$ are represented as
\bea
{\cal D}
\eq
\left\{
\vec{w}\in{\bf R}^N\left|
\vec{w}^{\rm T}\vec{e}=Nm,\vec{w}^{\rm T}\vec{w}\le N\tau
\right.
\right\},
\eea
and $\vec{e}=(1,1,\cdots,1)^{\rm T}\in{\bf R}^N$ is used.

As in previous work, 
in this research, we reformulate this stochastic optimization problem 
using the framework of statistical mechanical informatics to determine {the maximal} net present value per project 
$\kappa$ \cite{
1742-5468-2017-12-123402,
1742-5468-2016-12-123404,
doi:10.1080/1351847X.2011.601661,
KONDOR20071545,
PAFKA2003487,
Pafka2002,
Ciliberti2007,
doi:10.1080/14697680701422089,
110008689817,
doi:10.7566/JPSJ.86.063802,
doi:10.7566/JPSJ.86.124804,
SHINZATO2018986,
PhysRevE.94.052307,
PhysRevE.94.062102b,
10.1371/journal.pone.0134968,
10.1371/journal.pone.0133846,
1742-5468-2017-2-023301,
Ryosuke-Wakai2014}.
The partition function for this investment system at inverse temperature 
$\b$ is denoted by
\bea
Z(X,\b)
\eq\int_{\vec{w}\in{\cal D}}
d\vec{w}
%\exp\left({\b{\cal H}(\vec{w}|X)}\right)
e^{\b{\cal H}(\vec{w}|X)},
\eea
and then from 
\bea
\kappa\eq
\lim_{N\to\infty}
\f{1}{N}
\lim_{\b\to\infty}
\pp{}{\b}
\log Z(X,\b),
\label{eq8}
\eea
we can determine the {maximal} net present 
value per project $\kappa$.
Since we can evaluate $\kappa$ with the self-averaging property
using replica analysis{\cite{
110008689817,
doi:10.7566/JPSJ.86.063802,
doi:10.7566/JPSJ.86.124804,
SHINZATO2018986,
PhysRevE.94.052307,
PhysRevE.94.062102b,
10.1371/journal.pone.0134968,
10.1371/journal.pone.0133846,
1742-5468-2017-2-023301,
Ryosuke-Wakai2014}}, 
$E[Z^n(X,\b)]$, 
the $n$th moment of the partition function $Z(X,\b)$ is calculated as follows:
\bea
&&E[Z^n(X,\b)]\nn
\eq
\prod_{a}
\int_{\vec{w}_a\in{\cal D}}
d\vec{w}_a
E\left[
\exp\left(\b\sum_{a}{\cal H}(\vec{w}_a|X)\right)
\right]\nn
\eq
\f{1}{(2\pi)^{\f{Nn}{2}}}
\area
\prod_{i,a}dw_{ia}
E\left[
\exp
\left(
-\b\sum_{i,a}w_{ia}
\right.
\right.\nn
&&
+\b
\sum_{t=1}^T\f{1}{(1+r)^t}\sum_{i,a}c_iw_{ia}(1+x_{it})
\nn
&&+\b\sum_{i,a}\f{\l_iw_{ia}}{(1+r)^T}
+\sum_{a}k_a
\left(\sum_iw_{ia}-Nm\right)\nn
&&\left.\left.
-\f{1}{2}\sum_{a}\theta_a
\left(\sum_iw_{ia}^2-N\tau\right)
\right)\right],
\eea
where notations
$\sum_a=\sum_{a=1}^n$,
$\sum_{i}=\sum_{i=1}^N$,
$\prod_a=\prod_{a=1}^n$, and $\prod_i=\prod_{i=1}^N$ {are used}. 
Moreover, assuming there are a sufficiently large number of projects $N$, the average values with respect to stochastic fluctuations of the cash flow $x_{it}$ can be estimated as
\bea
&&
E\left[
\exp\left(
\b
\sum_{t=1}^T\f{1}{(1+r)^t}\sum_{i,a}c_iw_{ia}
(1+x_{it})
\right)
\right]\nn
\eq
\exp\left(
\b
\sum_{t=1}^T\f{1}{(1+r)^t}\sum_{i=1}^N
c_i\sum_{a=1}^nw_{ia}
\right.
\nn
&&
\left.
+\f{\b^2}{2}\sum_{t=1}^T\f{1}{(1+r)^{2t}}
\sum_{i=1}^Nc_i^2v_i
\left(\sum_{a=1}^nw_{ia}\right)^2
\right).
\qquad
\eea
This gives
\bea
&&
\log
E[Z^n(X,\b)]\nn
\eq
-Nn\b m
+\f{N\tau}{2}\sum_a\theta_a-Nm\sum_ak_a\nn
&&
+\log
\f{1}{(2\pi)^{\f{Nn}{2}}}
\area \prod_{i,a}dw_{ia}
\exp
\left(
-\f{1}{2}\sum_{a,i}\theta_aw_{ia}^2
\right.\nn
&&
+\sum_{i,a}w_{ia}\left(k_a+\b c_iA_1
+\f{\b\l_i}{(1+r)^T}
\right)\nn
&&
\left.
+\f{\b^2}{2}
A_2
\sum_ic_i^2v_i
\left(\sum_{a}w_{ia}\right)^2
\right)\nn
\eq
-Nn\b m+
\f{N\tau}{2}\sum_a\theta_a-Nm\sum_ak_a
\nn
&&
-\f{1}{2}
\sum_i
\log\det
\left|
\Theta_i
\right|\nn
&&
+\f{1}{2}
\sum_i
\left(\vec{k}+\b B_i\vec{e}_n\right)^{\rm T}
\Theta_i^{-1}%\nn
%&&
\left(\vec{k}+\b B_i\vec{e}_n\right),
\eea
where the constant vector $\vec{e}_n=(1,1,\cdots,1)^{\rm T}\in{\bf R}^n$, 
the vector of order parameters $\vec{k}=(k_1,k_2,\cdots,k_n)^{\rm T}\in{\bf R}^n$, 
$A_1=\sum_{t=1}^T\f{1}{(1+r)^{t}}=\f{1}{r}\left(1-\f{1}{(1+r)^T}\right)$,
$A_2=\sum_{t=1}^T\f{1}{(1+r)^{2t}}=\f{1}{r^2+2r}
\left(1-\f{1}{(1+r)^{2T}}\right)$, and 
$B_i=
c_iA_1+
\f{\l_i}{(1+r)^T}=
\f{c_i}{r}+\f{1}{(1+r)^T}\left(\l_i-\f{c_i}{r}\right)$
are defined.
\if 0
\bea
A_1\eq\sum_{t=1}^T\f{1}{(1+r)^{t}}\nn
\eq\f{1}{r}\left(1-\f{1}{(1+r)^T}\right),\\
A_2\eq\sum_{t=1}^T\f{1}{(1+r)^{2t}}\nn
\eq\f{1}{r^2+2r}
\left(1-\f{1}{(1+r)^{2T}}\right)
,\\
B_i\eq
c_iA_1+
\f{\l_i}{(1+r)^T}\nn
\eq
\f{c_i}{r}+\f{1}{(1+r)^T}\left(\l_i-\f{c_i}{r}\right).
\eea
\fi
Then, 
the element of the order parameter matrix 
$\Theta_i=
\left\{\theta_{i,ab}\right\}\in{\bf R}^{n\times n}$, $\theta_{i,ab}$, is defined as follows:
\bea
\theta_{i,ab}\eq
\left\{
\begin{array}{ll}
\theta_a-\b^2c_i^2v_iA_2&a=b\\
-\b^2c_i^2v_iA_2&a\ne b
\end{array}
\right..
\eea
As in previous work, 
the ansatz of the replica symmetry solution is used, that is, $k_a=k,\theta_a=\theta$, and
\bea
&&\lim_{N\to\infty}\f{1}{N}
\log E[Z^n(X,\b)]\nn
\eq-n\b m+\f{n\tau}{2}\theta
-nmk
-\f{1}{2}
\left\langle
\log
\left(\theta-n\b^2c^2vA_2\right)
\right\rangle\nn
&&-\f{n-1}{2}\log\theta
+\f{n}{2}
\left\langle
\f{\left(k+\b cA_1+\f{\b\l}{(1+r)^T}\right)^2}{\theta-n\b^2c^2vA_2}
\right\rangle,
\qquad
\eea
is analyzed. Here, 
we already use the notation 
\bea
\left\langle f(c,v,\l)
\right\rangle
\eq
\lim_{N\to\infty}
\f{1}{N}
\sum_{i=1}^Nf(c_i,v_i,\l_i).
\eea
In addition,
from the replica trick, 
\bea
\phi\eq
\lim_{N\to\infty}\f{1}{N}
E\left[\log Z(X,\b)\right]\nn
\eq
-\b m+
\f{\tau}{2}\theta-mk-\f{1}{2}\log\theta
+\f{\b^2A_2}{2\theta}\left\langle c^2v\right\rangle\nn
&&+\f{1}{2\theta}
\left\langle
\left(k+\b cA_1+\f{\b\l}{(1+r)^T}\right)^2
\right\rangle
\eea
is obtained. 
From the extremum of $k,\theta$, 
in the limit of the large inverse temperature $\b$,
\bea
k\eq-\b \left\langle 
cA_1+\f{\l}{(1+r)^T}
\right\rangle
+m\theta,\\
\f{\b}{\theta}
\eq\f{\sqrt{\tau-m^2}}
{\sqrt{A_2
\left\langle
c^2v
\right\rangle
+V
}},
\eea
are obtained, where
\bea
V
\eq
\left\langle
\left(A_1(c-
\left\langle
c
\right\rangle
)
+\f{\l-
\left\langle
\l
\right\rangle
}{(1+r)^T}
\right)^2
\right\rangle
\eea
is employed.
Using the alternative formulation in \siki{eq8}, 
we assess 
$\pp{\phi}{\b}=
-m+
\f{\b}{\theta}
A_2
\left\langle
c^2v
\right\rangle
+\f{k}{\theta}
\left\langle
\left(cA_1+\f{\l}{(1+r)^T}\right)
\right\rangle
+\f{\b}{\theta}
\left\langle
\left(cA_1+\f{\l}{(1+r)^T}\right)^2
\right\rangle$, 
and the maximal net present value per project is 
then
\bea
\kappa
\eq
\lim_{\b\to\infty}
\pp{\phi}{\b}\nn
\eq
-m
+mA_1\left\langle
c\right\rangle
+\f{m\left\langle
\l
\right\rangle}{(1+r)^T}
\nn
&&+\sqrt{\tau-m^2}
\sqrt{A_2
\left\langle
c^2v
\right\rangle
+V
}.\label{eq22}
\eea
Here, 
we can rewrite 
the investment amount ${w}_i$ in 
\siki{eq3}
as $w_i=mz_i$, where $m$ denotes the initial budget; then 
Eqs. (\ref{eq3}) and 
(\ref{eq4})
can be substituted with $\sum_{i=1}^Nz_i=N,\sum_{i=1}^Nz_i^2\le N\left(\f{\tau}{m^2}\right)$.
From the above, we can determine
the novel investment concentration, which is normalized by $m$, $\tau'=\f{\tau}{m^2}$. Then, 
\bea
\kappa
\eq
m\left(
-
+A_1\left\langle
c\right\rangle
+\f{\left\langle
\l
\right\rangle}{(1+r)^T}
\right.
\nn
&&
\left.
+\sqrt{\tau'-1}
\sqrt{A_2
\left\langle
c^2v
\right\rangle
+V
}
\right)
.
\eea
It turns out that the maximal net present value per project $\kappa$ is proportional to the initial budget $m$.
In this expression for the maximal net present value per project, 
by replacing
$w_i$ in \siki{eq1} with
$mz_i$, 
we can see that
${\rm NPV}_i$ and 
${\cal H}(\vec{w}|X)$
are 
proportional to $m$.

Next, we estimate the maximal net present value using an alternative method. The Lagrange multiplier function for maximization of the net present value, with budget and investment constraints, is defined as follows:
\bea
{\cal L}(\vec{w},k,\theta)
\eq{\cal H}(\vec{w}|X)
+k\left(\sum_{i=1}^Nw_i-Nm\right)\nn
&&
-\f{\theta}{2}
\left(\sum_{i=1}^Nw_i^2-N\tau\right),
\eea
where $k,\theta$ are the parameters representing the two constraints. From the extremum of the Lagrange multiplier function ${\cal L}(\vec{w},k,\theta)$, 
$\pp{{\cal L}}{w_i}=\pp{{\cal L}}{k}
=\pp{{\cal L}}{\theta}=0
$, the maximal net present value per project 
$\kappa$ is calculated, assuming the number of projects $N$ is sufficiently large:
\bea
\kappa
\eq
\lim_{N\to\infty}
\f{1}{N}\mathop{\max}_{\vec{w}\in{\cal D}}
{\cal H}(\vec{w}|X)\nn
\eq
m\left\langle 
h
\right\rangle
+\sqrt{\tau-m^2}
\sqrt{
\left\langle 
h^2
\right\rangle
-\left\langle 
h
\right\rangle^2
}\nn
\eq-m
+mA_1\left\langle
c\right\rangle
+\f{m\left\langle
\l
\right\rangle}{(1+r)^T}
\nn
&&+\sqrt{\tau-m^2}
\sqrt{A_2
\left\langle
c^2v
\right\rangle
+V
},\label{eq25}
\eea
where $h_i=-1+c_i\sum_{t=1}^T\f{1+x_{it}}{(1+r)^t}+\f{\l_i}{(1+r)^T}$, 
$\left\langle
h
\right\rangle=
\lim_{N\to\infty}
\f{1}{N}
\sum_{i=1}^Nh_i
=-1+A_1\left\langle c
\right\rangle
+\f{\left\langle \l
\right\rangle}{(1+r)^T}
$, 
$\left\langle h^2
\right\rangle-\left\langle h
\right\rangle^2=
\lim_{N\to\infty}\f{1}{N}
\sum_{i=1}^Nh_i^2-
\left\langle h
\right\rangle^2
=A_2
\left\langle
c^2v
\right\rangle
+V
$, 
$\theta=\sqrt{
\f{\left\langle h^2
\right\rangle-\left\langle h
\right\rangle^2}{\tau-m^2}
}$,
and 
$\f{k}{\theta}=m-\f{\left\langle h
\right\rangle}{\theta}$
{are obtained.}
From this, the optimal investment amount for project $i$ is 
$w_i=\f{k+h_i}{\theta}$.
Comparing the results of {\sikis{eq22}{eq25},} 
it turns out that 
the {theoretical} values obtained with replica analysis 
are consistent with those obtained with the Lagrange multiplier method. 

\begin{figure}[t] 
\begin{center}
\includegraphics[width=0.9\hsize]{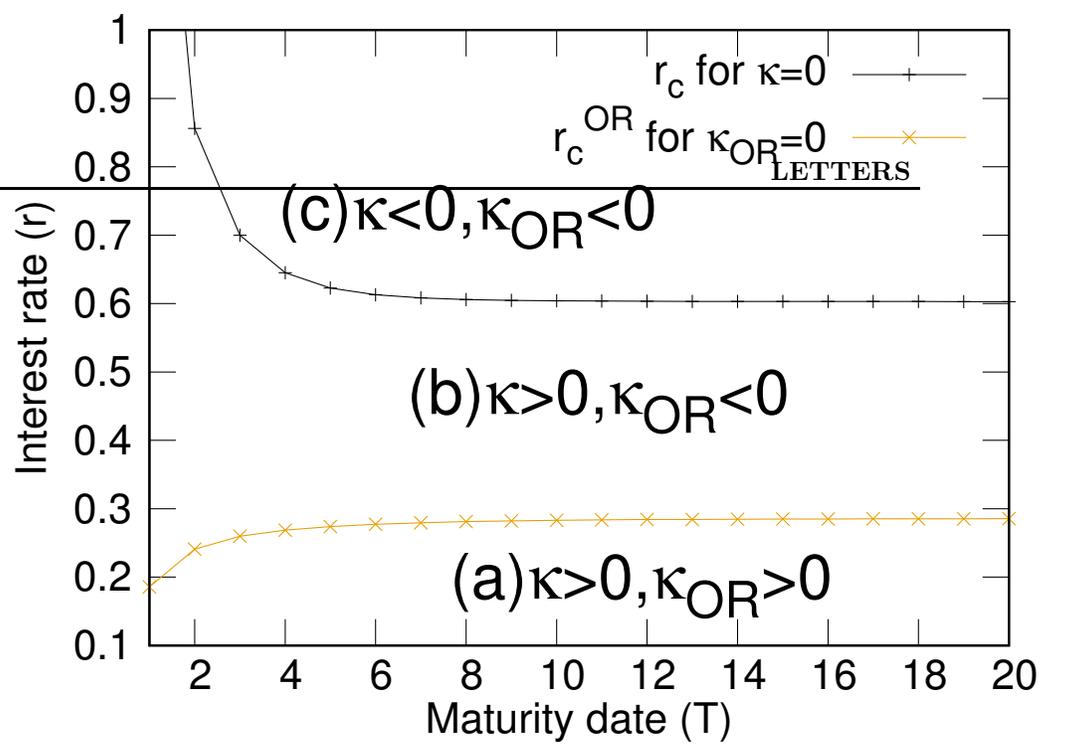}
\caption{
\label{Fig1}
Internal interest rate of the maximal net present value $r_c$ and 
internal interest rate of the maximal expected net present value $r_c^{\rm OR}$ versus maturity date $T$.
(a) region where $\kappa>0,\kappa_{\rm OR}>0$,
(b) region where $\kappa>0,\kappa_{\rm OR}<0$, and 
(c) region where $\kappa<0,\kappa_{\rm OR}<0$.
}
\end{center}
\end{figure}

{While in} the analytical approach based on operations research {
one first assesses the average of the
net present values, $E[{\cal H}(\vec{w}|X)]$, 
we maximize the expected net present value.} Previous studies have shown that the above operations research approach
is related to the analytic procedure for annealed disordered systems in spin glass theory. Thus, we can use the approach for quenched disordered systems to evaluate investment strategies under {both} constraints in this paper. 
Here, we analyze the maximal expected net present value per project ${\kappa}_{\rm OR}$, which is analogous to an annealed disordered system:
\bea
{\kappa}_{\rm OR}\eq\lim_{N\to\infty}\f{1}{N}\mathop{\max}_{\vec{w}\in{\cal D}}E[{\cal H}(\vec{w}|X)]\nn
\eq
-m
+mA_1\left\langle
c\right\rangle
+\f{m\left\langle
\l
\right\rangle}{(1+r)^T}
,\label{eq23}
\eea
where the maximal net present value per project $\kappa$ in {\sikis{eq22}{eq25}} and 
the maximal expected net present value 
$\kappa_{\rm OR}$ 
in \siki{eq23}
are compared. As a result, 
\bea
\kappa\ge\kappa_{\rm OR}.
\eea
 Moreover, the maximal net present value $\kappa$ is related to the 
variance of the stochastic fluctuations of the cash flow, $V[x_{it}]=v_i$, that is, $\left\langle c^2v\right\rangle$, 
though the maximal expected net present value $\kappa_{\rm OR}$
does not depended on the variance of $x_{it}$. 
With
our proposed approach, by treating 
{the net present value} as a quenched disordered system, it is possible to evaluate the influence of stochastic fluctuations in the cash flow 
on $\kappa$, which is not possible with 
the analytical procedure for annealed disordered systems.

In general, we can compare and contrast investment portfolios based on their net present values. In particular, the interest rate $r$, which is called the internal interest rate when the net present value is equal to zero, is a useful indicator of how profitable an investment is. Here,
when the distributions of the parameters $c_i,\l_i,v_i$ are known, 
we can evaluate 
the internal interest rate $r_c$ for the maximal net present value $\kappa$ and 
the internal interest rate $r_c^{\rm OR}$ for the maximal expected net present value $\kappa_{\rm OR}$.
The coupon rate $c_i$ of project $i$ is i.i.d. and follows a beta distribution (shape parameters $\a,\b(>0)$, and density function $f_c(c_i)=\f{c_i^{\a-1}(1-c_i)^{\b-1}}{B(\a,\b)},0<c_i<1$), and the 
attenuation rate of the sales amount for project $i$, $\l_i$, is also i.i.d. and follows an exponential distribution (with mean $\g$ and density function $f_\l(\l_i)=\f{1}{\g}e^{-\f{\l_i}{\g}},0<\l_i$);
furthermore, the variance of the stochastic fluctuation of the cash flow $x_{it}$, $v_i$, is 1.
Here, when $\a=2,\b=5,\g=0.9,\tau'=\f{\tau}{m^2}=3$, 
the internal interest rates $r_c,r_c^{\rm OR}$ as a function of maturity date $T$ are as shown in Figure \ref{Fig1}.
When the interest rate $r$ is under both lines, 
the maximal net present value $\kappa$ and 
the maximal expected net present value $\kappa_{\rm OR}$ are both positive, and when 
the interest rate $r$ is above both lines, 
the maximal net present value $\kappa$ and 
the maximal expected net present value $\kappa_{\rm OR}$ are negative.
{For regions (a) and (c) in the figure,
since the signs of $\kappa$ and 
 $\kappa_{\rm OR}$ are consistent with each other, investment judgment based on the results of 
the operations research approach is not misled.
In contrast, since in the region (b) 
 the maximal net present value $\kappa$ is positive and the maximal expected net present value $\kappa_{\rm OR}$ is negative, that is, they are not consistent,
investment judgment based on the maximal expected net present value $\kappa_{\rm OR}$ creates the possibility of an investment loss because of $\kappa>0$ holding in region (b).}

%\section{}
In this work, we have considered the use of methods for analyzing quenched {disordered} systems for optimizing the net present value of investment portfolios containing multiple projects.
Using the framework of statistical mechanical informatics, we represented {the object function of the investment system as a Hamiltonian,} which is defined by the sum of the net present value of each project,
and we succeeded in deriving an expression for the maximal net present value. We then derived an expression for the maximal expected net present value using the analytical approach developed in operations research (the analytical procedure for annealed {disordered} systems), 
and {we verified that the maximal net present value is always larger than the maximal expected net present value because an approach commonly used in operations research underestimates the true maximal net present value.
Furthermore, 
through a numerical experiment, we revealed that 
the regions of positive and negative {maximal} expected net present values obtained by the conventional analytic method in operations research are not consistent with the other theoretical bounds.
}
From this, 
by including a stochastic component in the expression for the net present value, the optimal investment portfolio can be estimated with greater accuracy.

In this paper, 
we have assumed that $t=0$ is the time at which all projects are invested in simultaneously, 
and the maturity date $T$ and interest rate $r$ are the same for all projects. 
However, these assumptions are not realistic, and will be generalized in future work. 
Furthermore, our method needs to allow not only the investment amount but also the timing of the investment to vary by project.

%\section*{Acknowledgements}
The author is grateful for detailed discussions with D. Tada and H. Yamamoto. 
This work was supported in part by
Grants-in-Aid Nos. 15K20999, 17K01260, and 17K01249; the Research Project of the Institute of Economic Research Foundation at Kyoto University; and Research Project
No. 4 of the Kampo Foundation.

%\appendix
%\section{Moments\label{app-a}}

%\bibliographystyle{plain}
%\bibliographystyle{unsrt}
%\bibliographystyle{apsrev4-1}
%\bibliographystyle{amsplain}
\bibliographystyle{jpsj}
\bibliography{sample20180913}

\end{document}